\documentclass[preprint]{aastex6}
\usepackage{xspace}
\usepackage{color}
\definecolor{DarkGreen}{rgb}{0.0,0.45,0.0}  

\newcommand{\kms}{km~s$^{-1}$\xspace}

\newcommand{\rsun}{$R_\odot$\xspace}

\shorttitle{Wave Impact on Coronal Structures}
\shortauthors{Liu et al.}

\begin{document}

\title{Impacts of EUV Wavefronts on Coronal Structures in Homologous Coronal Mass Ejections}

\author{Rui Liu\altaffilmark{1}, Yuming Wang\altaffilmark{1}, Jeongwoo Lee\altaffilmark{2}, Chenglong Shen\altaffilmark{1}}
\altaffiltext{1}{CAS Key Laboratory of Geospace Environment, Department of Geophysics and Planetary Sciences, University of Science and Technology of China, Hefei 230026, China; rliu@ustc.edu.cn}
\altaffiltext{2}{Institute of Space Sciences, Shandong University, Weihai 264209, China}
\begin{abstract}
Large-scale propagating fronts are frequently observed during solar eruptions, yet it is open whether they are waves or not, partly because the propagation is modulated by coronal structures, whose magnetic field we still cannot measure. However, when a front impacts coronal structures, an opportunity arises for us to look into the magnetic properties of both interacting parties in the low-$\beta$ corona. Here we studied large-scale EUV fronts accompanying three coronal mass ejections (CMEs), each originating from a kinking rope-like structure in the NOAA active region (AR) 12371. These eruptions were homologous and the surrounding coronal structures remained stationary. Hence we treated the events as one observed from three different viewing angles, and found that the primary front directly associated with the CME consistently transmits through 1) a polar coronal hole, 2) the ends of a crescent-shaped equatorial coronal hole, leaving a stationary front outlining its AR-facing boundary, and 3) two quiescent filaments, producing slow and diffuse secondary fronts. The primary front also propagates along an arcade of coronal loops and slows down due to foreshortening at the far side, where local plasma heating is indicated by an enhancement in 211~{\AA} (\ion{Fe}{14}) but a dimming in 193~{\AA} (\ion{Fe}{12}) and 171~{\AA} (\ion{Fe}{9}). The strength of coronal magnetic field is therefore estimated to be $\sim\,$2~G in the polar coronal hole and $\sim\,$4~G in the coronal arcade neighboring the active region. These observations substantiate the wave nature of the primary front and shed new light on slow fronts. 

\end{abstract}

\section{Introduction}
Large-scale propagating fronts associated with solar flares and coronal mass ejections (CMEs) have been under intensive study for decades \citep[see][for recent reviews]{Patsourakos+Vourlidas2012,Liu+Ofman2014,Warmuth2015,Chen2016rev,Long2017}, owing mostly to three generations of space-borne telescopes with ever increasing spatiotemporal resolution in EUV, namely, the Extreme-ultraviolet Imaging Telescope \citep[EIT;][]{Delaboudiniere1995} onboard the Solar and Heliospheric Observatory \citep[SOHO;][]{Domingo1995}, the Extreme UltraViolet Imager \citep[EUVI;][]{Wuelser2004} onboard the Solar Terrestrial Relationals Observatory \cite[STEREO;][]{Kaiser2008}, and the Atmospheric Imaging Assembly \citep[AIA;][]{lemen12} onboard the Solar Dynamics Observatory \citep[SDO;][]{pesnell12}. Hence they are often referred to as ``EIT waves'' or ``EUV waves''. However, their physical nature is still under debate, as to whether they are fast MHD waves propagating in the corona or ``pseudo waves'' generated by magnetic restructuring associated with an expanding CME. There is evidence for a hybrid picture comprising an outer, fast-mode MHD wavefront and an inner, CME-associated non-wave front \citep[e.g.,][]{Liu+Ofman2014,Chen2016rev}. 

One of the most convincing arguments for the wave interpretation comes from the evidence for reflection and refraction at regions with strong gradients in Alfv\'{e}n and fast-magnetosonic speeds, typically at the boundary of ARs and coronal holes. It has long been noticed that EUV waves tend to avoid ARs and coronal holes \citep[refraction; e.g.,][]{Thompson2000}. The reflection of EUV waves at the coronal hole boundary has been reported in several cases  \citep{Gopa2009,Li2012,Olmedo2012,Shen+Liu2012,Shen2013,Yang2013}. The EUV wave transmission through a coronal hole is relatively rare \citep{Olmedo2012,LiuW2018}. In addition, \citet{Veronig2006} found that a Moreton wave slides into a coronal hole up to 100 Mm. Similarly, the transmission of coronal waves into ARs is obscure and rare. The wavefront becomes very faint within ARs, only re-emerging from the far side \citep{Li2012,Shen2013}, which is understood by the conservation of wave energy flux. Sometimes coronal waves reflect at ARs \citep{Shen2013,Kumar+Manoharan2013}. Occasionally, secondary wavefronts are produced when the primary wavefront impacts on coronal loops \citep{Kumar+Manoharan2013}. 

It has long been known that coronal waves can cause the `winking' of filaments, i.e., a filament fades or disappears and then re-appears in the H$\alpha$ line center due to wave-triggered oscillations \citep[e.g.,][]{rs66,Liu2013}. Such oscillations are relatively rare and have much larger amplitude \citep[$>20$ km s${}^{-1}$; see the review by][]{tripathi09} than the frequently observed small-amplitude oscillations \citep[$\sim 2-3$ km s${}^{-1}$; see the review by][]{arregui12}, the latter of which are usually local and seemingly intrinsic. \citet{LiuW2012} reported the transmission of an EUV wave through a coronal cavity with enhanced speed, causing coherent oscillations of filament threads embedded in the cavity. It is puzzling that a filament does not always oscillate when a wave passes by \citep{okamoto04,Liu2013}. Apparently, the filament's height and magnetic environment as well as its orientation with respect to the wavefront are significant factors deciding how it responds to the wave passage \citep{Liu2013,Shen2014filament,Zhang2016}. Investigations on such interactions could yield important insight into physical properties of both interacting parties \citep[e.g.,][]{Gilbert2008,LiuW2012}.

Here we present observations of large-scale EUV fronts associated with a series of halo CMEs originating from the same AR, NOAA 12371, which are conventionally termed (quasi-)homologous CMEs \citep{Liu2017,Lugaz2017}. These events provide a precious opportunity to study interactions of the fronts with various coronal structures from different viewing angles, given that the waves are homologous and propagating in similar coronal environments. The structures impacted by fronts include a polar coronal hole, an equatorial coronal hole, two quiescent filaments, and a coronal arcade neighboring to the source AR. In the following text, all the observed propagating fronts are referred to as wavefronts for simplicity. The detailed analysis of the observations is presented in \S2;  interpretations and implications of the observations are discussed and summarized in \S3.

\section{Observation \& Analysis}

AR 12371 produced four halo CMEs during its transit on the solar disk from 2015 June 16--28, each associated with an M-class flare, a large-scale EUV wave, and a metric Type II radio burst observed by the WAVES instruments on-board Wind and STEREO spacecrafts (not shown). However, the wave associated with the CME on June 18 failed to make discernible effects on coronal structures on the disk, while other coronal structures of interest were still behind the limb. We hence focus on the EUV waves associated with the later three CMEs on 2015 June 21, 22 and 25, respectively (Figure~\ref{fig:ltc}). The June 22 event occurred when AR 12371 was located near the disk center, and hence is investigated in detail (\S\ref{subsec:0622}). Results from this disk-center event are corroborated by the other two events providing complementary viewing angles (\S\ref{subsec:0621} and \S\ref{subsec:0625}). 

In the investigations below we mainly used the EUV imaging data obtained by AIA. The instrument takes full-disk images with a spatial scale of 0.6 arcsec pixel$^{-1}$ and a cadence of 12 s. Among the 7 EUV and 2 UV passbands, we focused on four of them: 131~{\AA} (primarily \ion{Fe}{21} for flare plasma, with a peak response temperature $\log T = 7.05$; \ion{Fe}{8} for ARs, $\log T = 5.6$), 211~{\AA} (\ion{Fe}{14}, $\log T = 6.3$), 193~{\AA} (\ion{Fe}{24} for flare plasma, $\log T = 7.25$; \ion{Fe}{12} for ARs, $\log T = 6.2$) and 171~{\AA} (\ion{Fe}{9}, $\log T = 5.8$). The 131~{\AA} passband is preferentially used to detect hot eruptive structures, while the other three passbands are ideal for the detection of EUV wavefronts. The flares were also observed in hard X-rays (HXRs) by the Reuven Ramaty High-Energy Solar Spectroscopic Imager \citep[RHESSI;][]{lin02} and the Gamma-ray Burst Monitor (GBM) of the Fermi Gamma-ray Space Telescope. The CMEs were observed by the Large Angle and Spectrometric Coronagraph Experiment (LASCO) on-board the Solar and Heliospheric Observatory (SOHO).

\subsection{2015 June 22 Event} \label{subsec:0622}

\subsubsection{Eruption \& Wave Initiation}
\begin{figure} 
	\centering
	\includegraphics[width=0.85\hsize]{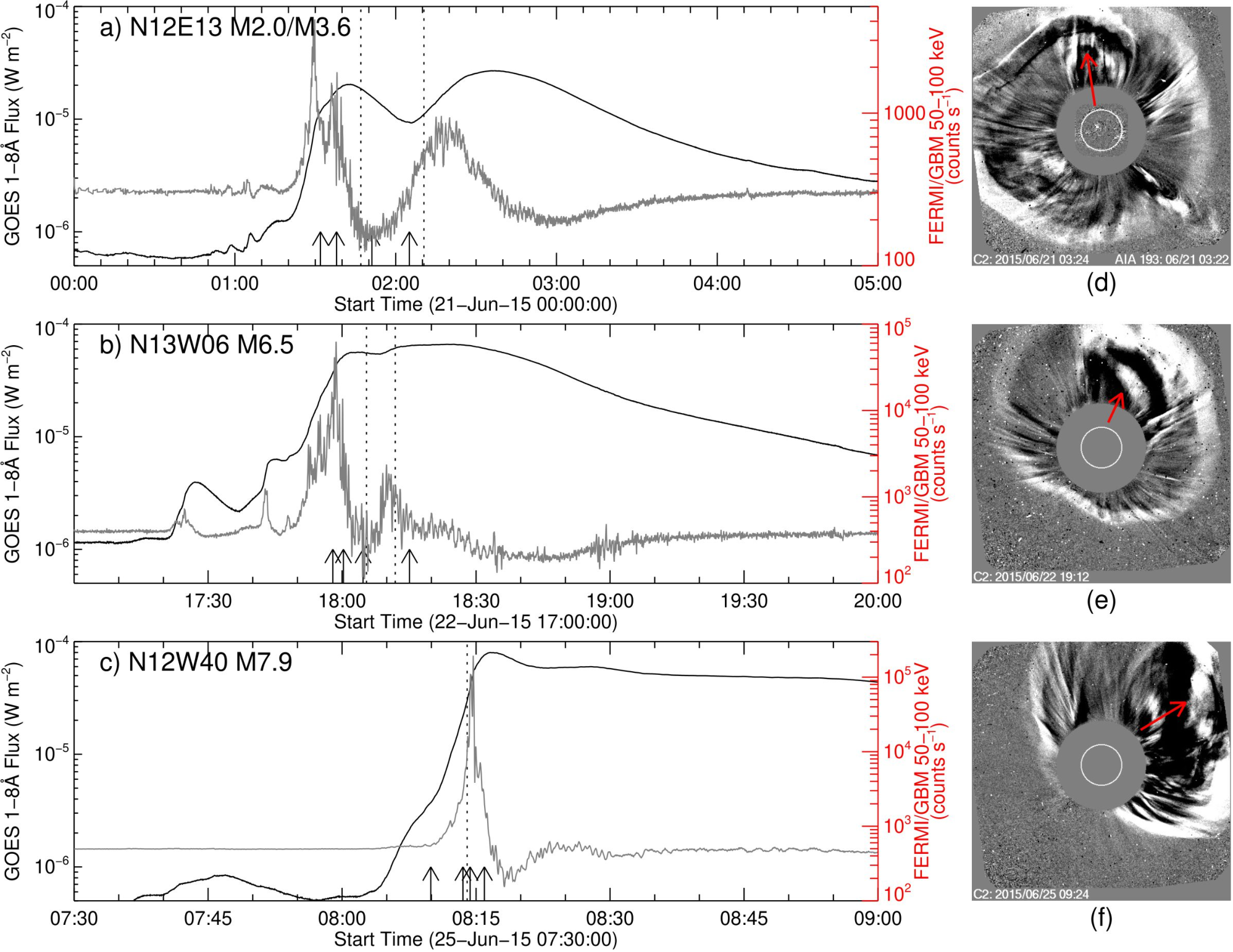}
	\caption{Homologous CMEs and accompanying flares from AR 12371. \emph{Left column} shows the flare lightcurves: GOES 1--8~{\AA} fluxes are scaled by the left $y$-axis, its time derivative (gray) is shown in an arbitrary unit; HXR count rates at 50--100 keV recorded by Fermi/GBM are scaled by the right $y$-axis (red). The GBM missed the first HXR burst at about 01:30 UT on 2015 June 21. The dotted line marks when a large-scale EUV wavefront was first detected. Note two successive wavefronts were seen on 2015 June 21 and 22. Black arrows at the bottom of each panel mark the times of the AIA images in Figures~\ref{fig:init}, \ref{fig:init0621}, and \ref{fig:init0625}.  Flare locations in heliographic coordinates are indicated in each panel. \emph{Right column} shows the halo CMEs observed by the C2 camera of LASCO. Red arrows indicate the inner fronts and the expulsion direction observed in the low corona.  \label{fig:ltc}}
\end{figure}

\begin{figure} 
	\centering
	\includegraphics[width=0.85\hsize]{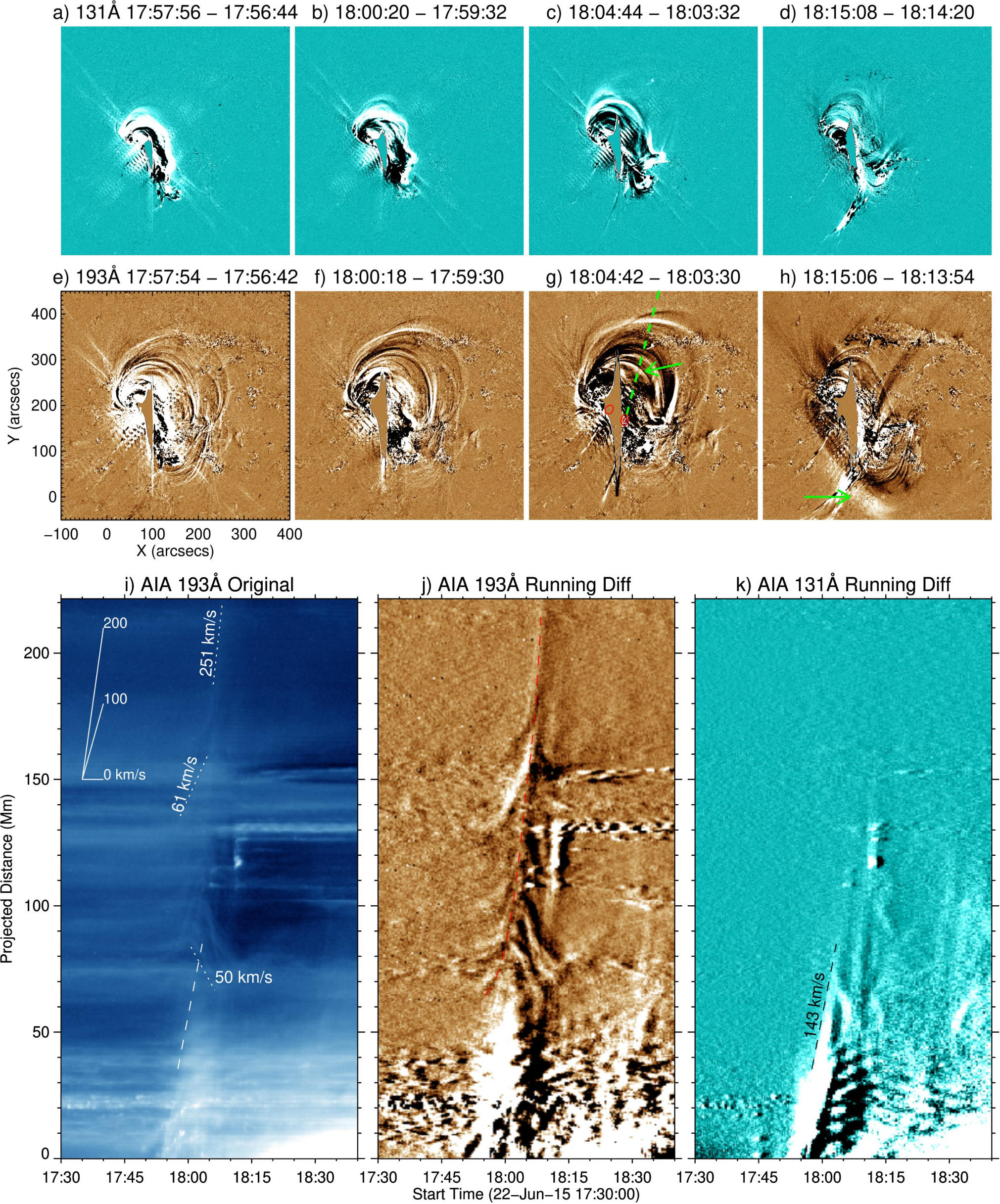}
	\caption{Initiation of the eruption in AR 12371 on 2015 June 22. (a--d) A rope-like structure (RLS) in running difference images of AIA 131~{\AA}. (e--h) Coronal loops overlying the RLS in AIA 193~{\AA}. The arrow in (g) marks a loop undergoing contraction. RHESSI HXR footpoints are shown as contours in (g). Panels (i--k) show the dynamics seen through the virtual slit in (h), using original images in AIA 193~{\AA}, running difference images in AIA 193~{\AA} and 131~{\AA}, respectively. Dotted lines in (i) indicate the linear fitting to various features, including loop expansion (61 \kms), loop contraction (50 \kms), wavefront propagation (251 \kms). The dashed line fitting the rising RLS in (k) is replotted in (i). An animation of 131 and 193~\AA running difference images is available online. \label{fig:init}}
\end{figure}

The M6.5-class flare on 2015 June 22 has been studied from various perspectives, e.g., flare precursors on the surface \citep{Wang2017} and in the corona \citep{Awasthi2018}, fine structures and loop slippage during the decay phase \citep{Jing2016,Jing2017}. Here we concentrate on the eruption initiation and EUV waves. At about 17:49 UT, a low-lying hot loop under expansion appeared in AIA 131~{\AA} in the center of AR 12371. Starting from about 17:58 UT (Figure~\ref{fig:init}a), which was close to the 1st time-derivative peak of GOES 1-8~{\AA} flux (Figure~\ref{fig:ltc}b), the expanding loop appeared to be increasingly twisted/writhed with time (Figure~\ref{fig:init}(b--c)), which we term a rope-like structure (RLS). The RLS corresponds to a complex flux-rope system revealed by nonlinear force-free field models \citep{Awasthi2018}. The overlying coronal loops in AIA 193~{\AA} first expanded and then contracted (Figure~\ref{fig:init}(e--g)). In difference images, an expanding loop has a bright outer rim and a dark inner rim, while the reverse is true for a contracting loop. Located right above the RLS, an exemplary loop under contraction is marked by an arrow in Figure~\ref{fig:init}g. 

A virtual slit (marked by a green dashed line in Figure~\ref{fig:init}g) is placed across the AR loops and the RLS, and a time-distance map is generated by stacking up the slices taken from images in the chronological order. A northward-propagating wavefront was detached from the AR loops at about 18:06 UT at 250 \kms on the time-distance map (Figure~\ref{fig:init}(i--j)). The wavefront can be traced back into the AR via a continuous half-bell-shaped track on the time-distance map (delineated by a red dashed curve in Figure~\ref{fig:init}j), with short stripes veering on both sides: loop expansion (contraction) produces positively (negatively) sloped stripes on the left (right), with comparable speeds of $\sim\,$50 \kms. The coronal loops right above the RLS started to contract at about 18:00 UT. The RLS rose and expanded at 140 \kms between about 17:57 and 18:03 UT (Figure~\ref{fig:init}k). Starting from about 18:12 UT, the time close to the 2nd SXR peak (Figure~\ref{fig:ltc}b), a wavefront emerged with a jet-like feature (marked by an arrow in Figure~\ref{fig:init}h, see also Figure~\ref{fig:init}d), propagated southward, and became diffused by 18:23 UT.  

\subsubsection{Wave Propagation}
\begin{figure} 
	\centering
	\includegraphics[width=0.9\hsize]{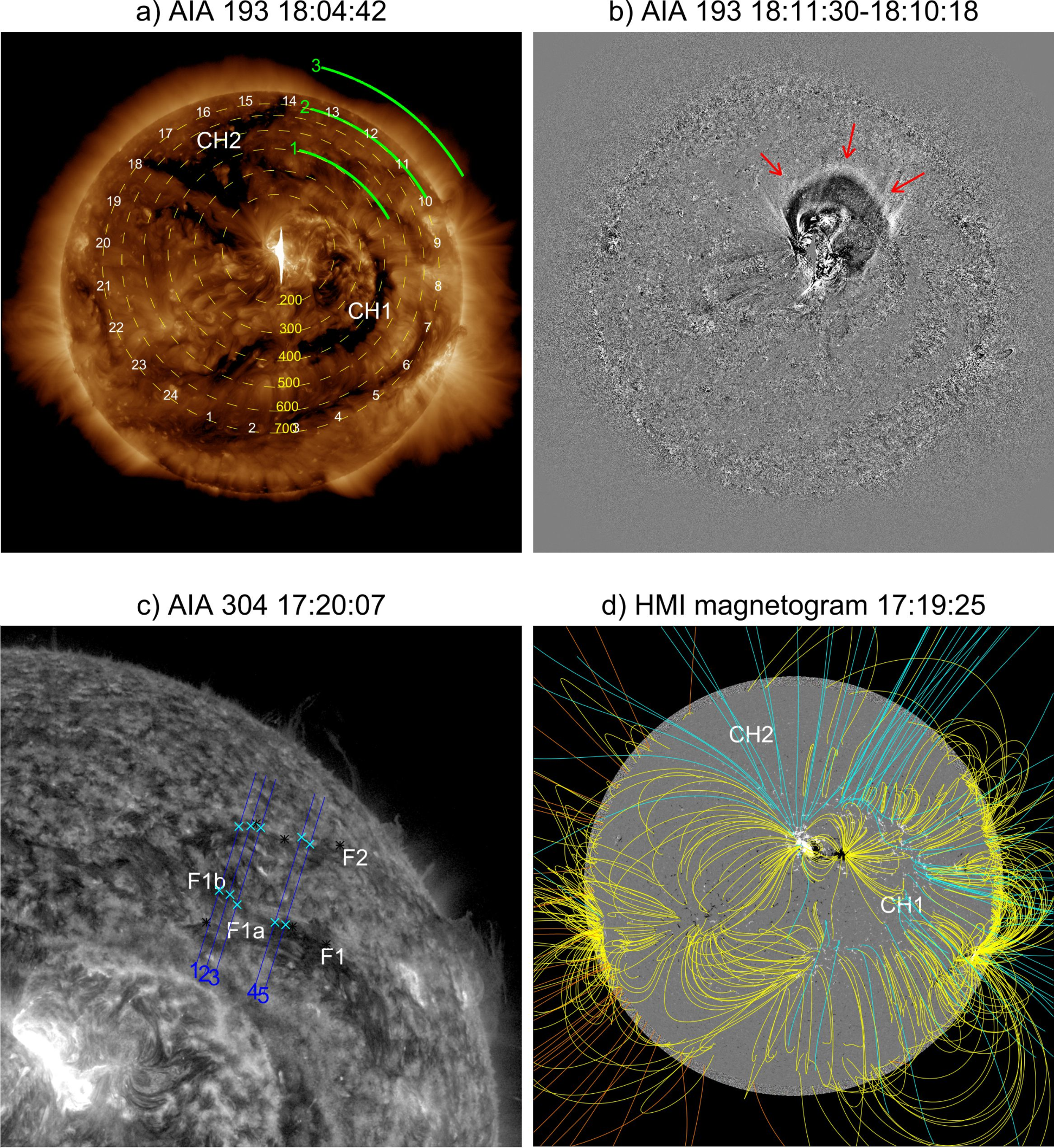} %
	\caption{Virtual slits used to study the wave propagation. a) 24 sector-shaped slices (white) centered on the flare and 3 representative arc-shaped slices (green) concentric to the disk center. Yellow dashed curves indicate the distance from the sector center, with numbers in units of Mm. b) The primary wavefront (marked by arrows) in an AIA 193~{\AA} difference image. c) Linear slices across the two quiescent filaments labeled F1 and F2 in an AIA 304 image. The eastern section of F1 bifurcates into two branches labeled F1a and F1b. The `x' symbols correspond to the cyan reference lines in Figure~\ref{fig:ff}. d) PFSS field lines are superposed on a line-of-sight magnetogram obtained by the Helioseismic and Magnetic Imager (HMI) onboard SDO. The magnetogram is saturated at $\pm200$ G. Yellow indicate closed field lines; cyan and orange indicate open field lines originating from positive and negative polarity, respectively. An animation of 193~{\AA} running difference images is available online. \label{fig:slit}}
\end{figure}

Unlike the jet-associated wavefront, the primary wavefront propagated mainly in the northwest direction (Figure~\ref{fig:slit}b). To study the wave propagation, we divided the solar disk into 24 sectors, each spanning 15 deg (Figure~\ref{fig:slit}a) and centering on the midpoint of the conjugated HXR footpoints at 25--50 keV at 18:04:40 UT (Figure~\ref{fig:init}g). Each sector-shaped slice is converted to a 1-dimensional slice by averaging over the azimuthal direction. Stacking up the slices chronologically yields the time-distance maps in Figure~\ref{fig:sector}. One can see that the jet-associated wavefront is detected in Sectors 1--4 (labeled `JWF'), propagating at about 300~\kms, while the primary wavefront, which was initiated earlier, is detected mainly in Sectors 8--17, propagating at a speed exceeding 700~\kms to as far as over 800 Mm away from the flaring site.  

The wave propagation seems to be modulated by the strength of the local field. Here we utilized a potential-field-source-surface \citep[PFSS;][]{Schrijver+Derosa2003} model to shed light on this matter. In Figure~\ref{fig:slit}d the starting points to trace field lines are randomly selected on the surface but weighted by magnetic flux so that there are \replaced{less number of}{fewer} field lines in regions of weaker field. One can see that the magnetic field is generally weak to the north of AR 12371, through which the primary wavefront propagated. Another weak-field region is located to the immediate south of AR 12371, which may explain the propagation of the jet-associated wavefront in this direction.

Coronal structures that were impacted by the wave include an equatorial coronal hole (CH1), a north polar coronal hole (CH2; Figure~\ref{fig:slit}(a \& d)), and two quiescent filaments (F1 and F2; Figure~\ref{fig:slit}c). Note the western end of F2 can be seen above the limb, embedded at the bottom of a coronal cavity (Figure~\ref{fig:slit}a). Below we will investigate in detail the impact of the EUV wave on the aforementioned coronal structures. 

\subsubsection{Wave Impact on Coronal Structures}

\begin{figure} 
	\centering
	\includegraphics[width=\hsize]{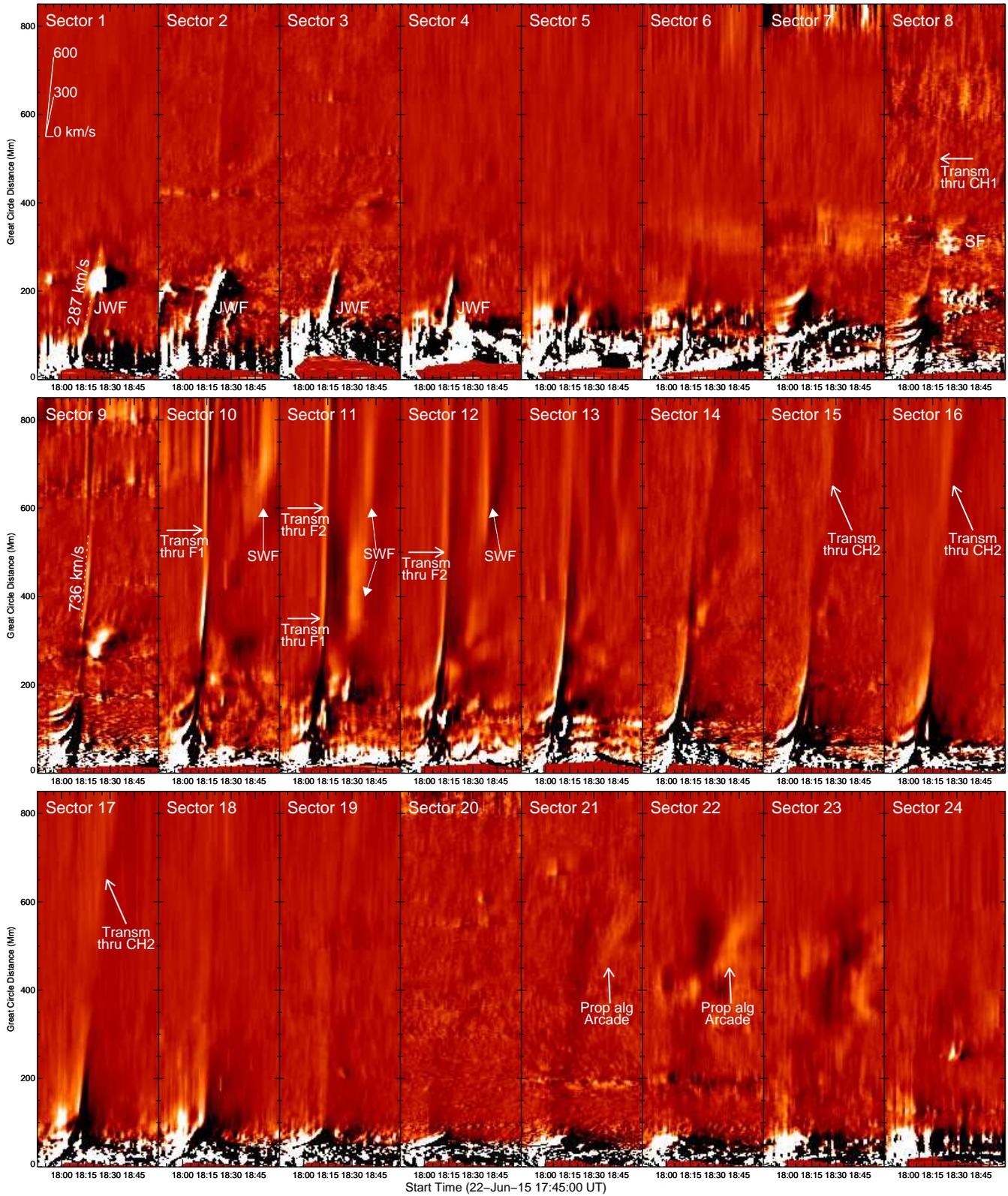}
	\caption{Time-distance maps constructed via sector-shaped slices in Figure~\ref{fig:slit}a, using AIA 193~{\AA} running difference images. The jet-associated wavefront is labeled FWF in Sectors 1--4. The primary wavefront is labeled PWF. Secondary wavefronts produced when the primary wavefront impacting filaments are labeled SWF. \label{fig:sector}}
\end{figure}

\begin{figure} 
	\centering
	\includegraphics[width=\hsize]{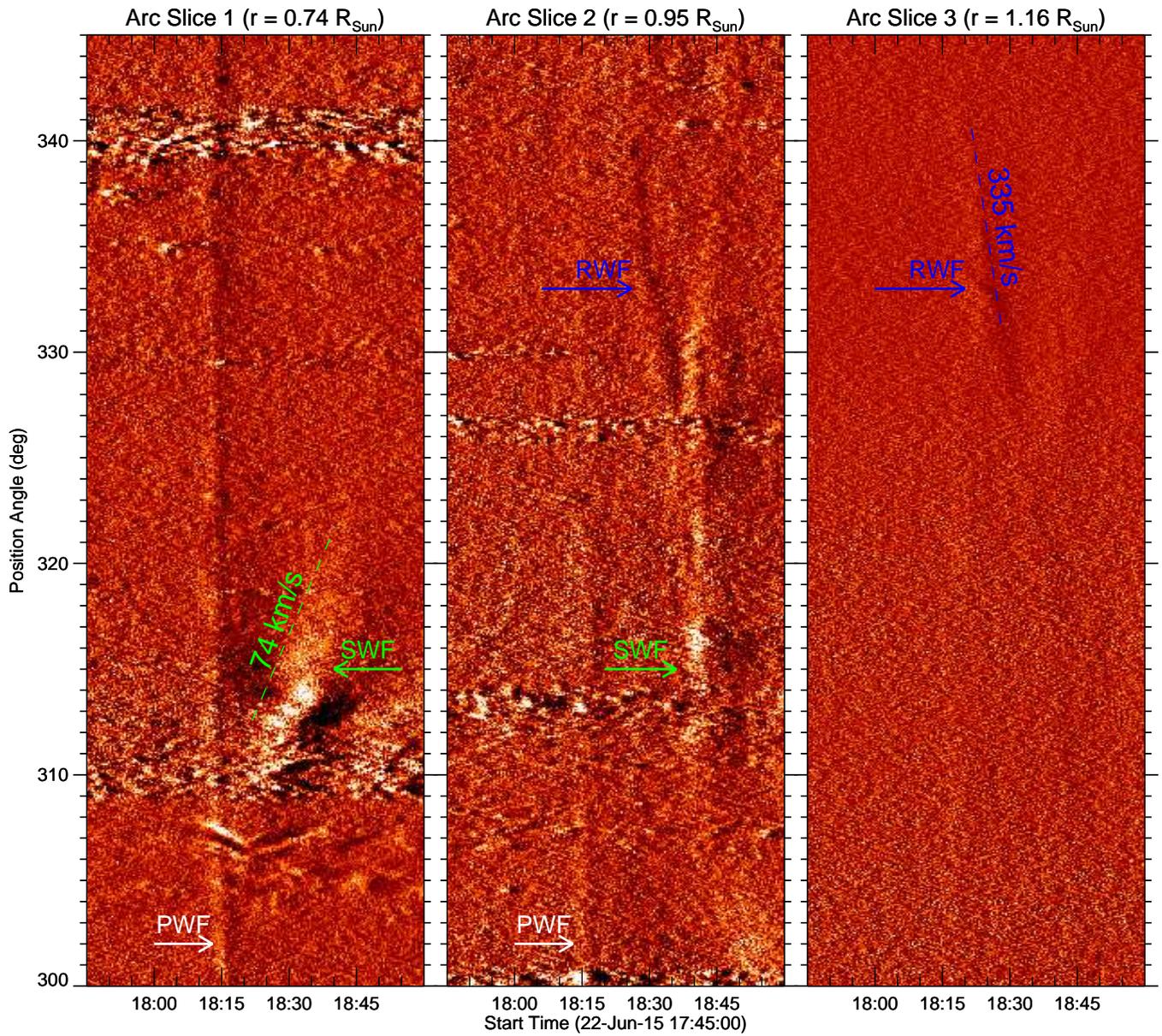}
	\caption{Time-distance maps constructed via arcs concentric to the disk center in Figure~\ref{fig:slit}a, using AIA 193~{\AA} running difference images. The primary wavefront is labeled PWF, the wavefront reflected/refracted from the coronal hole CH2 is labeled RWF, and secondary wavefronts induced at the filaments are labeled SWF.\label{fig:arc}}
\end{figure}

\paragraph{Wave \& Coronal Holes} The wave impact on coronal holes can be seen through Sectors 4--8, which cover CH1, and Sectors 15--17, which cover CH2. In Figure~\ref{fig:sector}, the wave transmission through CH1 is only detected in Sector 8, which covers the northern end of the crescent-shaped CH1, but not in Sectors 4-7. In contrast, the wave transmission through CH2 is detected in all sectors across it. Note a stationary front was produced at the AR-facing boundary of CH1 (labeled `SF' in Figure~\ref{fig:prop}(d--f); see also Figure~\ref{fig:sector}, Sector 8). In addition, a wavefront moving away from CH2 towards the equator can be seen in the polar region above the limb in AIA 193~{\AA} running difference images (see the animation accompanying Figure~\ref{fig:slit}). It was also detected from the arc slits close to as well as above the limb up to 0.16 \rsun ($\sim\,$110 Mm), starting at about 18:21 UT at $PA\approx340$ deg with an apparent speed of over 300 \kms (labeled `RWF' in Figure~\ref{fig:arc}). It soon became very diffuse at $PA \approx 330$ deg, before being able to impact F2 at $PA\approx310$. The wavefront must be either reflected off or refracted out of CH2 since it propagated away from CH2 and appeared later than the arrival of the primary wavefront (labeled `PWF') at about 18:15 UT. 

\paragraph{Wave \& Coronal Arcade}

The wave transmission through an arcade of coronal loops was most clearly visible in 211~{\AA} (see also the animation accompanying Figure~\ref{fig:prop}), but also marginally visible in 193~{\AA} (Sectors 21--23 in Figure~\ref{fig:sector}). This arcade consists of coronal loops connecting positive flux in the eastern AR and negative flux to the east of AR (Figure~\ref{fig:slit}d and Figure~\ref{fig:prop}(a--c)). The wavefront appears to propagate along these loops (indicated by red arrows in Figure~\ref{fig:prop}(d--f)), initially at $\sim\,$600 \kms (Figure~\ref{fig:prop}g), but then significantly decelerated to $\sim\,$50 \kms as it propagated toward the far (eastern) side of the arcade. The wavefront was enhanced in 211~{\AA} (Figure~\ref{fig:prop}g) but dimmed in 193 and 171~{\AA} (Figure~\ref{fig:prop}(h and i)), especially when it approached the eastern end of the arcade, where the propagation apparently stopped.

\begin{figure} 
	\centering
	\includegraphics[width=\hsize]{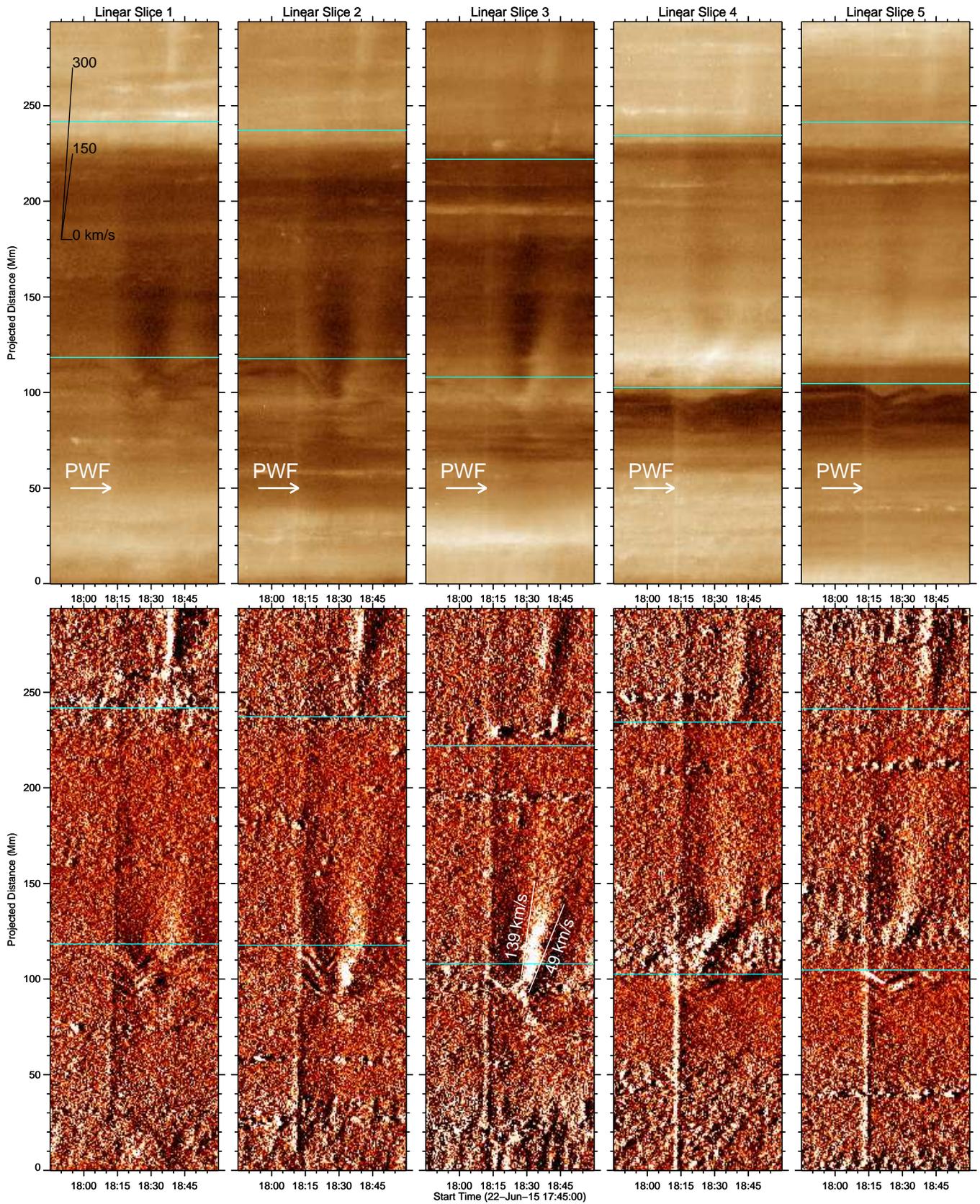}
	\caption{Time-distance maps constructed via linear slices in Figure~\ref{fig:slit}(c). The cyan lines correspond to the reference points (crosses) in Figure~\ref{fig:slit}(c) \label{fig:ff}}
\end{figure}

\paragraph{Wave \& Filaments}
The wave transmission through the filaments F1 and F2 is detected in Sectors 10--12 (Figure~\ref{fig:sector}). There is no discernible change of the primary wavefront as it traversed the filaments. In addition to the fast and sharp primary wavefront, one or two secondary wavefronts that are slow and diffuse are also visible in the corresponding time-distance maps. The secondary wavefronts are closely related with the filament disturbance in response to the impact of the primary wavefront, but this information is lost in the average over the azimuthal direction in each sector. We hence constructed time-distance maps (Figure~\ref{fig:ff}) with linear slices across the two filaments (Figure~\ref{fig:slit}c). We picked some reference points on the northern edge of the filaments along the slices (cyan crosses; Figure~\ref{fig:slit}c), each corresponding to a horizontal reference line in the time-distance map (Figure~\ref{fig:ff}). Note F1's eastern section bifurcated, labeled F1a and F1b. Under the impact of the primary wavefront, F1's displacement as large as 5--10 Mm can be seen in the time-distance maps southward of the reference line at around $y=110$ Mm. F2's displacement is not quite visible, due to poorer contrast and more severe foreshortening nearer the limb. With the reference points we found that secondary wavefronts originated from the northern side of F1 and F2. These wavefronts initiated either from F1 soon after the impact of the primary wavefront at about 18:15 UT (Linear Slices 4 and 5; Figure~\ref{fig:ff}), or when the displaced filament swung back, i.e., after the half period of the oscillation (Linear Slices 1--3). They are very diffuse and hence can be detected in various slits oriented in different directions, including the arc-shaped slices (Figure~\ref{fig:arc}(a \& b)). The speed was estimated to range from tens of kilometers per second to about 150 \kms.

\subsection{2015 June 21 Event} \label{subsec:0621}
\subsubsection{Eruption \& Wave Initiation}
\begin{figure} 
	\centering
	\includegraphics[width=0.85\hsize]{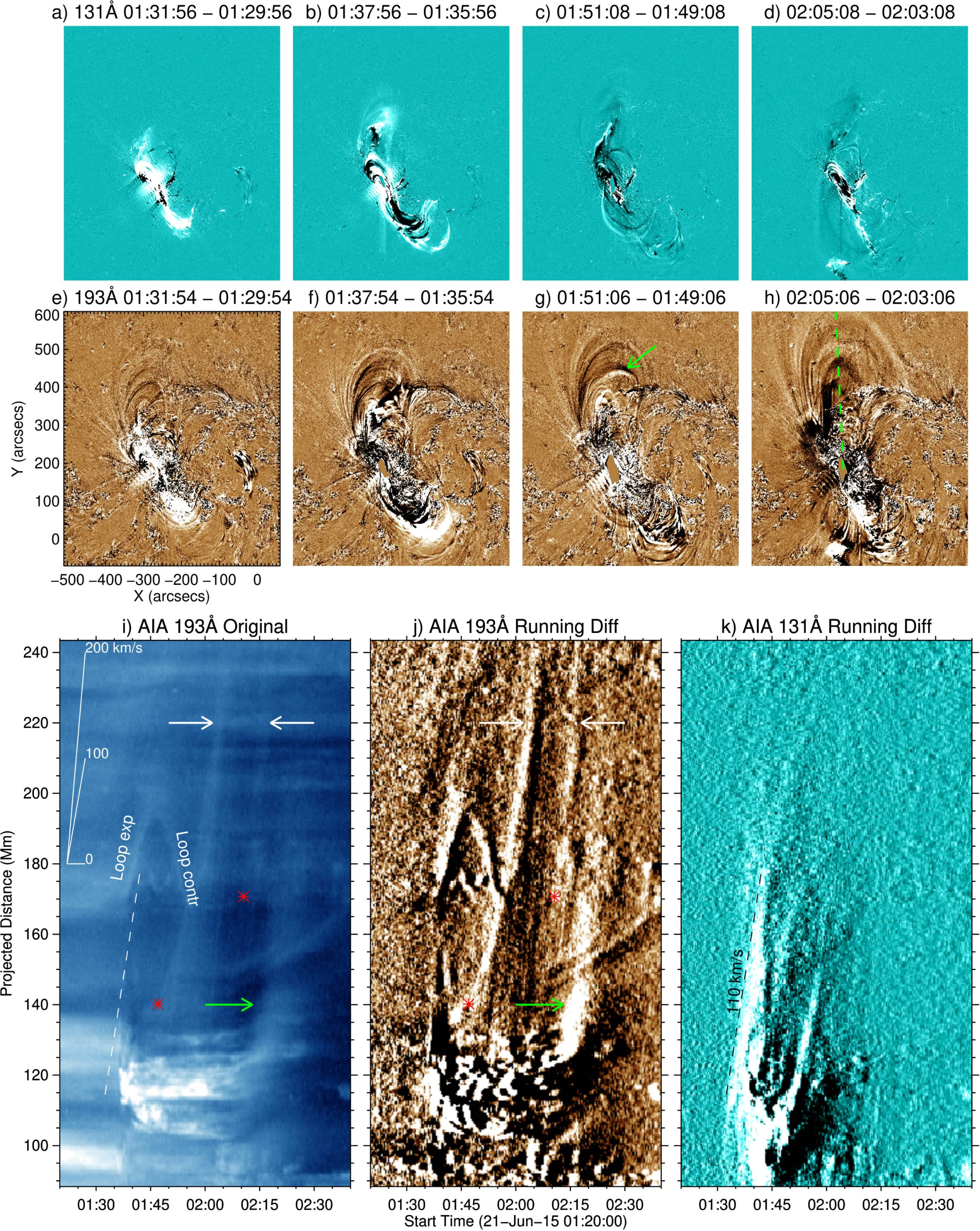}
	\caption{Initiation of the eruption in AR 12371 on 2015 June 21. (a--d) A rope-like structure (RLS) in running difference 131~{\AA} images. (e--h) Coronal loops overlying the RLS in running difference 193~{\AA} images. The arrow in (g) marks a representative loop undergoing contraction. Panels (i--k) show the dynamics seen through the virtual slit in (h), using original  193~{\AA} images, running difference 193~{\AA} and 131~{\AA} images, respectively. The dashed line fitting the rising RLS in (k) is replotted in (i). In (i and j), the white arrows mark the two successive wavefronts, whose initiating time and location along the slit are marked by red asterisks; the green arrow marks a jet-like feature (see also Figure~\ref{fig:wave0621}d). An animation of 131 and 193~{\AA} running difference images is available online. \label{fig:init0621}}
\end{figure}

On 2015 June 21, the halo CME was associated with two successive M-class flares, M2.0 and M3.6 from AR 12371, which peaked in SXRs at 01:42 and 02:36 UT, respectively (Figure~\ref{fig:ltc}b). A sigmoidal RLS was observed in 131~{\AA} \citep[Figure~\ref{fig:init0621}a; see also][]{Lee2018}) during the rising phase of the M2.0 flare. This structure was apparently kinked and slightly rotated clockwise as it rose and expanded (Figure~\ref{fig:init0621}(b--c)). Meanwhile in 193~{\AA}, a bundle of higher coronal loops in the north first expanded and then contracted (Figure~\ref{fig:init0621}(e--g)). A representative loop undergoing contraction is marked by an arrow in Figure~\ref{fig:init0621}g. Through a virtual slit starting from the flaring site and oriented in the expanding direction of the coronal loops (dashed line in Figure~\ref{fig:init0621}h), one can see in the resultant time-distance maps (Figure~\ref{fig:init0621}(i--k)) that the deflection of coronal loops during 01:37--01:54 UT, i.e., expansion followed by contraction, was closely associated with the ascent of the RLS at about 110~\kms. Right after the loop expansion, an EUV (primary) wavefront was observed to initiate at 01:47 UT at about 140 Mm from the flaring site (Figure~\ref{fig:init0621}(i--j)). A wavefront immediately following the primary one initiated at 02:10 UT at about 170 Mm from the flaring site (Figure~\ref{fig:init0621}(i--j)), which was apparently associated with a jet-like feature (marked by green arrows in Figures~\ref{fig:init0621}(i--j) and \ref{fig:wave0621}d). Both wavefronts leave a smooth continuous track on the time-distance maps made from 193~{\AA} images. 

\subsubsection{Wave Propagation \& Impact}
\begin{figure} 
	\centering
	\includegraphics[width=\hsize]{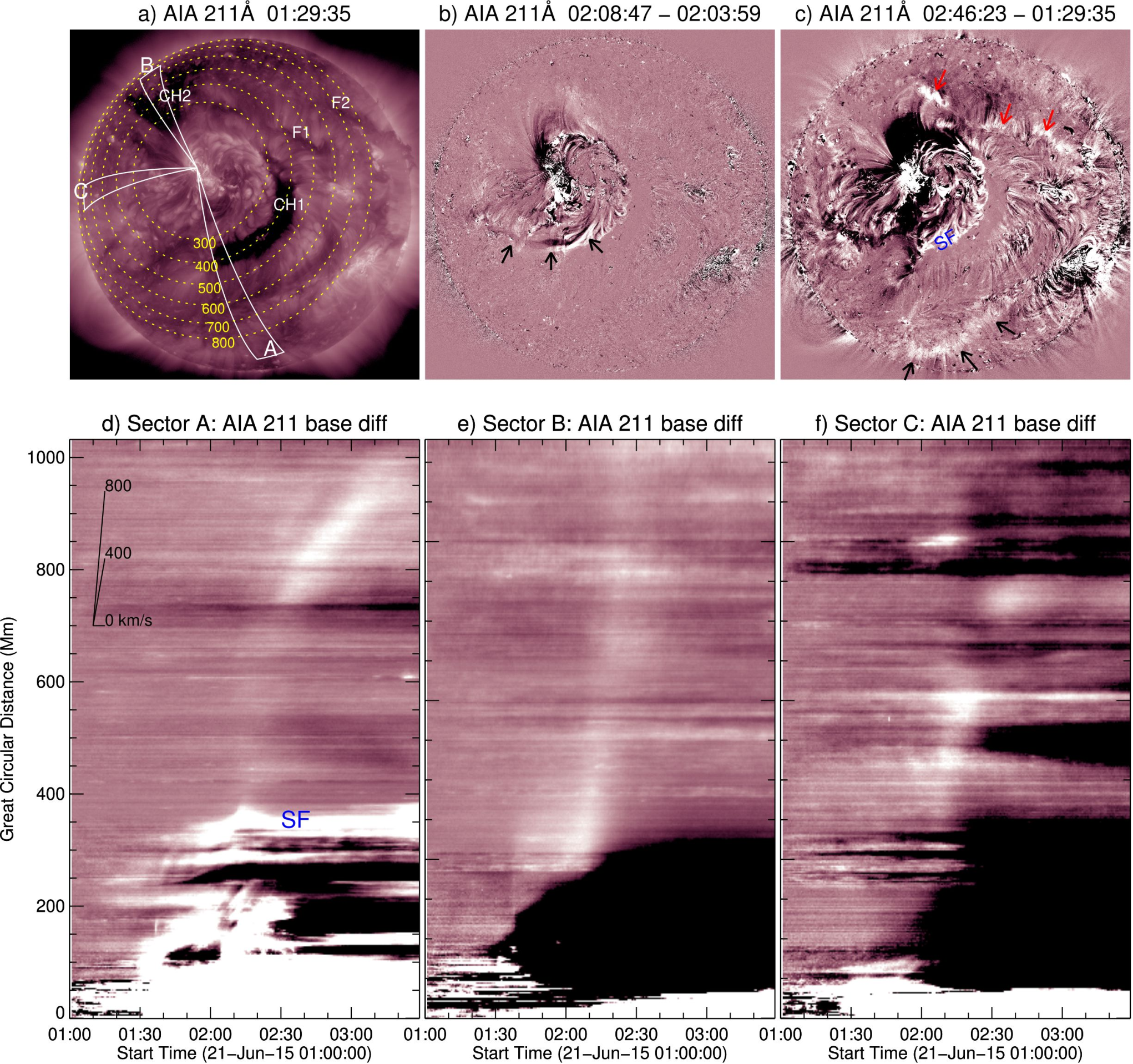}
	\caption{Wave propagation in the 2015 June 21 event. Panel (a) shows three representative sector-shaped slices (labeled A, B, and C). Yellow dashed circles indicate the distance from the sector center, i.e., the flaring site, with numbers in units of Mm. Panels (b \& c) highlight in 211~{\AA} difference images the southward-propagating wavefront (black arrows), the secondary wavefronts produced at the filament F1 (red arrows), and the stationary front (labeled SF) at the eastern boundary of CH1. Time-distance maps constructed via the sector-shaped slices in (a) show the wave transmission through CH1 (d) and CH2 (e), and wave propagation along the arcade neighboring the active region (f). An animation of 211 and 193~{\AA} running difference images is available online. \label{fig:wave0621}}
\end{figure}

To study the wave propagation, we again employed 24 sectors to cover the solar disk (Figure~\ref{fig:wave0621}(a)), each spanning 15 deg and centering on the midpoint of the conjugated HXR footpoints at 25--50 keV at around 01:26:30 UT (not shown). Three representative sectors are shown in Figure~\ref{fig:wave0621}(a). The primary wavefront propagated mainly southward and northward. The southward-propagating wavefront (marked by black arrows in Figure~\ref{fig:wave0621}(b \& c)) transmitted through CH1 along Sector A (Figure~\ref{fig:wave0621}d), but not in other angular directions. The wavefront was weak inside CH1, and a stationary front formed at the AR-facing boundary of CH1, taking a similar crescent shape as CH1 (Figures~\ref{fig:wave0621}f).  The northward-propagating wavefront transmitted through CH2 along Sector B (Figure~\ref{fig:wave0621}e). Unlike CH1, the transmission is seen in various angular directions covering the whole coronal hole, without a stationary front at the boundary. Along Sector C (Figure~\ref{fig:wave0621}f), the transmission through the arcade to the east of AR 12371 is only marginally visible due to its proximity to the limb.

\begin{figure} 
	\centering
	\includegraphics[width=0.9\hsize]{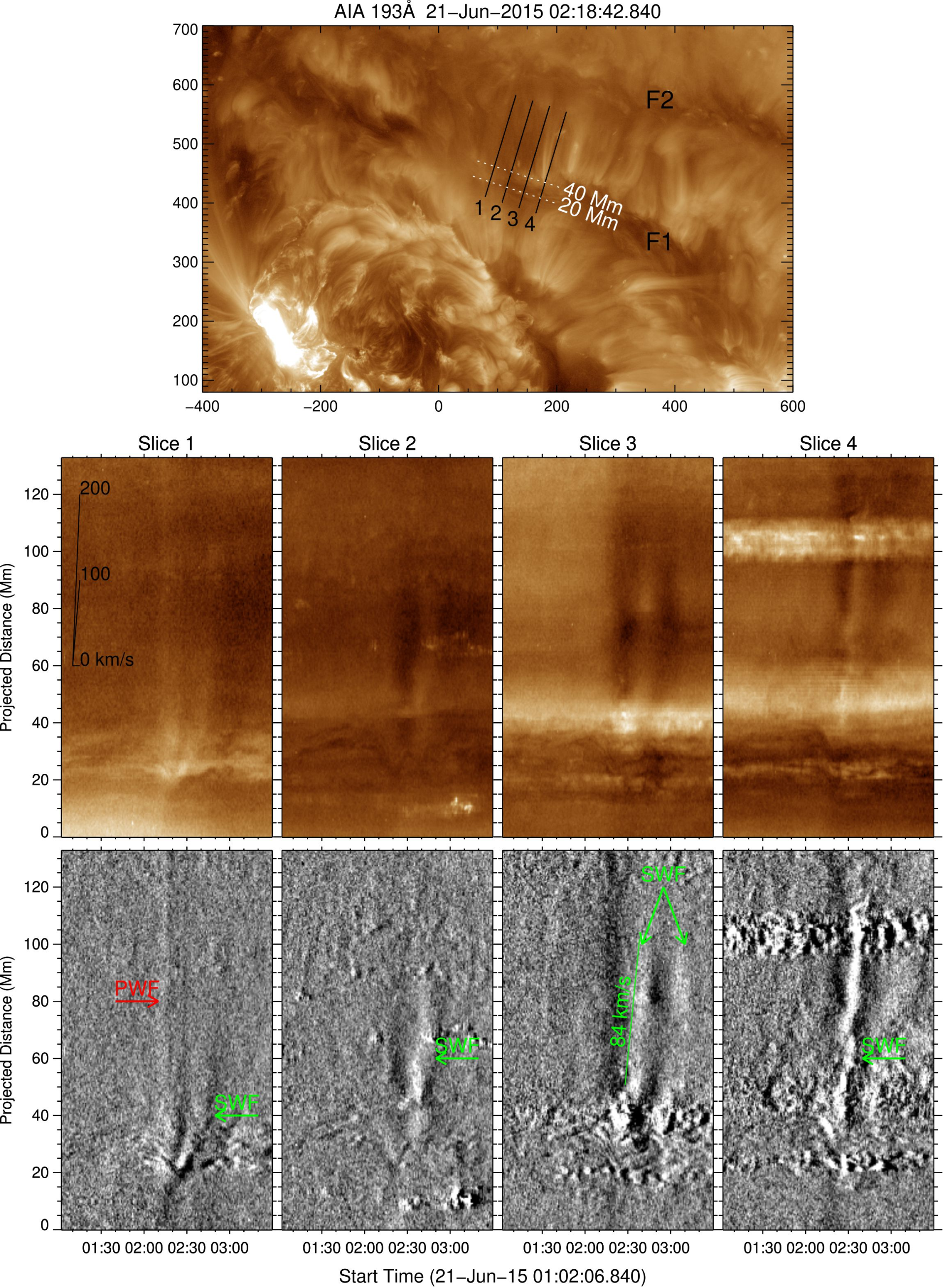}
	\caption{Secondary wavefronts generated at the filament F1. Panel a) shows the four linear slices across F1. Middle and bottom panels show the time-distance maps constructed via the linear slices, using original and running difference 193~{\AA} images, respectively. \label{fig:ff0621}}
\end{figure}

Double or single secondary wavefronts can again be seen through virtual slices across the filament F1 (Figure~\ref{fig:ff0621}). All the secondary wavefronts were associated with the filament disturbances between 20--40 Mm along the slices (marked by dotted lines in the top panel of Figure~\ref{fig:ff0621}) at speeds of no more than 100 \kms. The filament disturbances are better discernible in time-distance maps constructed using original (middle panels) than running difference images (bottom panels); the reverse is true for the secondary wavefronts. 

\subsection{2015 June 25 Event} \label{subsec:0625}
\begin{figure} 
	\centering
	\includegraphics[width=0.85\hsize]{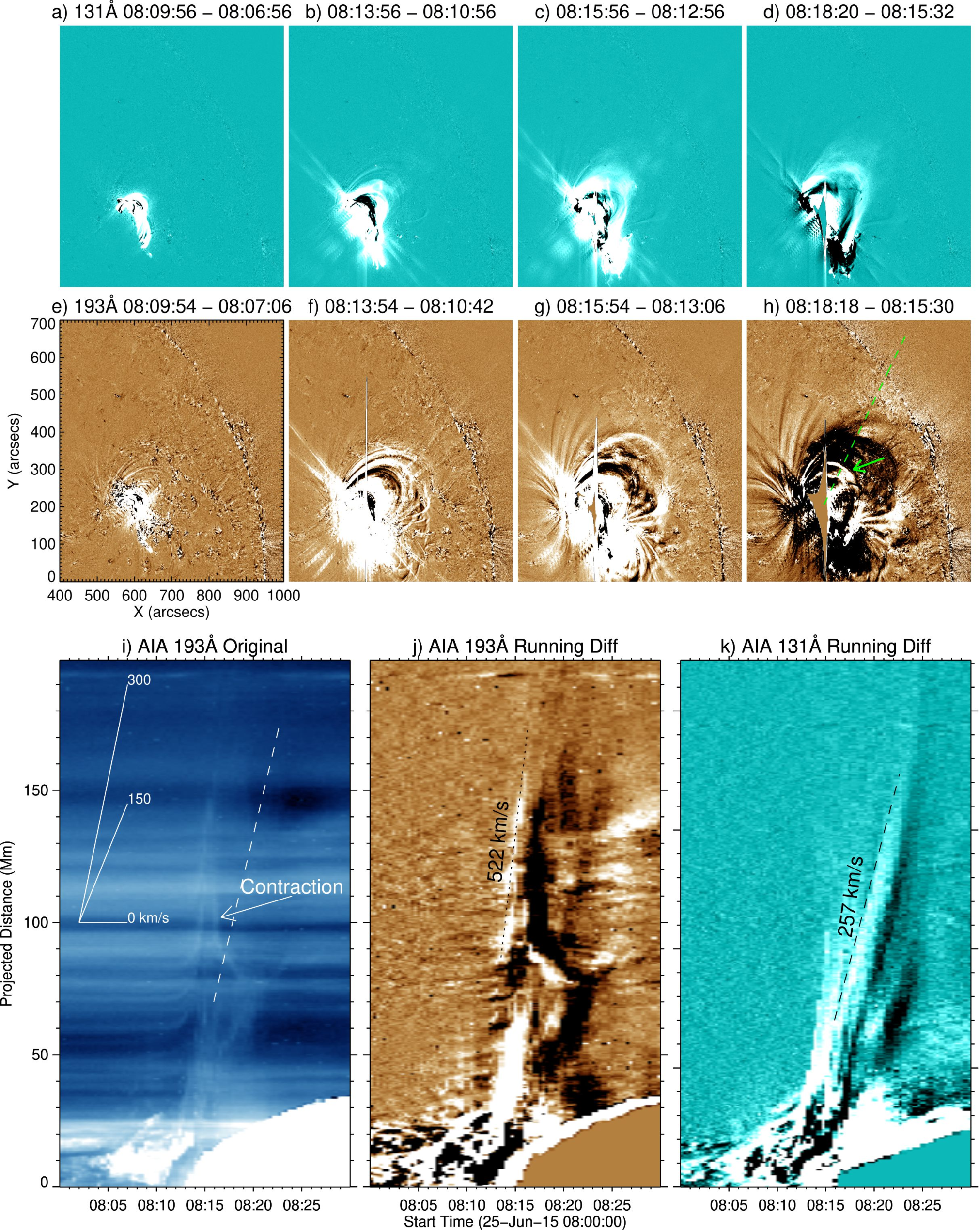}
	\caption{Initiation of the eruption in AR 12371 on 2015 June 25. Top Panels (a--d) show a rope-like structure (RLS) in 131~{\AA} running difference images. Middle Panels (e--h) show coronal loops overlying the RLS in 193~{\AA} running difference images. The arrow in (h) marks a representative loop undergoing contraction. Bottom Panels (i--k) show the dynamics seen through the virtual slit in (h), using original images in 193~{\AA}, running difference images in 193 and 131~{\AA}, respectively. In (j) the wavefront is fitted by a dotted line. The dashed line fitting the rising RLS in (k) is replotted in (i). An animation of 131 and 193~{\AA} running difference images is available online. \label{fig:init0625}}
\end{figure}

At the onset of the M7.9 flare at about 08:10 UT on 2015 June 25, one can see two groups of sheared loops in 131~{\AA}, apparently crossing each other (Fig.~\ref{fig:init0625}a). These loops soon evolved into an eruptive RLS, reminiscent of tether-cutting reconnection \cite{Liu2010}. The RLS' southern leg was apparently kinked (Fig.~\ref{fig:init0625}(b \& c)). The primary wavefront emerged at about 08:15 UT, taking a loop shape (Figure~\ref{fig:init0625}g). From a virtual slit oriented along the expulsion direction, one can see that the wave initiation was again associated with the expansion and subsequent contraction of coronal loops overlying the RLS (Fig.~\ref{fig:init0625}(i \& j)). A representative 193~{\AA} loop under contraction is marked by an arrow in Fig.~\ref{fig:init0625}h. The wave initiation time was around the HXR peak at 50--100 keV at 08:14:28 UT (Figure~\ref{fig:ltc}d). 

In this event, the wave impact on coronal structures is best seen for CH2 and the arcade to the east of AR 12371 (Figure~\ref{fig:wave0625}a). CH1 was located too close to the western limb, despite that a stationary front again formed at its AR-facing boundary (Figure~\ref{fig:wave0625}b). The primary wavefront also passed through filament F1 and produced a secondary wavefront propagating northward (Figures~\ref{fig:wave0625}b; see also the animation accompanying Figure~\ref{fig:wave0625}). With sector-shaped slices centered on the flaring site, one can see through Sector A that the primary wavefront transmitted \replaced{trough}{through} CH2 (Fig.~\ref{fig:wave0625}(c,d--f)). Though weaker inside CH2 than outside, generally the wavefront was slightly enhanced in 211~{\AA}, significantly enhanced in 193~{\AA}, but dimmed in 171~{\AA}, suggesting that the plasma \replaced{a}{at} the wavefront was warmed up to about 1.5 MK, an effect not clearly seen in the other two events. Sector B highlights the wavefront propagation along the arcade of coronal loops to the east of AR 12371, with an apparent deceleration (Figure~\ref{fig:wave0625}(g--i)). The wavefront was initially shown as an enhanced feature in 211 and 193~{\AA}, but not quite visible in 171~{\AA}. However, when approaching the far side of the arcade, the wavefront was transformed into a dimmed feature in 171 and 193~\replaced{{AA}}{{\AA}}, while remained enhanced in 211~{\AA}, suggesting that the plasma at the wavefront was warmed up to about 2 MK, similar to the June 22 event (Figure~\ref{fig:prop}(g--i)).

\section{Discussion \& Conclusion}

\subsection{Homologousness of the Events}
AR 12371 dominated the northern hemisphere during its disk transit. During the same period AR 12367 was the only major active region in the southern hemisphere; it was far away from AR 12371 and already close to the western limb on 2015 June 21 (c.f. Figure~\ref{fig:wave0621}(a \& c)). Thus the EUV waves under investigation were propagating in similar coronal environments, interacting with similar coronal structures. This allows us to study these interactions from different viewing angles. On the other hand, the waves also derive their similarities from the homologousness of the eruptions, which are elaborated below from three aspects. 

First, the eruptions produced CMEs with similar morphology. All three CMEs exhibit a two-front morphology \citep{Vourlidas2013}: a halo outer front followed by an inner front with a more limited angular width (Figure~\ref{fig:ltc}(d--f)). The former is likely to be caused by the density compression at the primary wavefront, while the latter is identified with the RLS observed in the low corona, judging by its expulsion direction  and spatial extent. Whether or not the wavefront is driven by the CME is debatable because of the asymmetry between the CME front with a limited angular width and the halo shocked front. \citet{Howard+Pizzo2016} argued that the asymmetry could be explained by a blast wave and a flux rope erupting together. The contrast between a halo wavefront in the extended corona and a much narrower EUV wavefront on the surface suggests that the wave propagates in all directions in the relatively homogeneous high corona but is confined in the ``valleys'' of low Alf\'{v}en speed in the highly inhomogeneous low corona.

Second, the eruptions originate from the same segment of the AR's polarity inversion line, shared similar eruptive structure, and exhibited similar dynamics in the active region. The RLS in each event appeared to be kinked before erupting into higher corona (Figures~\ref{fig:init}, \ref{fig:init0621}, \ref{fig:init0625}), hence we interpret it as a magnetic flux rope. The writhing motion indicates that the rope's magnetic twist has reached the threshold of the helical kink instability \citep{Gilbert2007}. The increase in twist can be accomplished by reconnections during the early phase of the flare \citep[e.g.,][]{Wang2017NC}. Moreover, when the kinking RLS ascended, the overlying loops deflected, i.e., first expanded and then contracted (Figures~\ref{fig:init}, \ref{fig:init0621}, and \ref{fig:init0625}). A similar phenomenon was reported when a kinking filament `squeezed' through the overlying arcade \citep{Liu+Wang2009}.  Alternatively, the loop deflection is expected with the passage of a compressive wavefront followed by a rarefaction in pressure. This might be true in the June 22 and 25 events, in which the wavefront track in the time-distance map passes apparently through the `watershed' between loop expansion and contraction (Figures~\ref{fig:init} \& \ref{fig:init0625}), but is not the case in the June 21 event (Figure~\ref{fig:init0621}). Hence we lean toward the former scenario.

Third, all three events show two major peaks in SXR or its time derivative, indicating two episodes of energy release (Figure~\ref{fig:ltc}(a--c)). In the June 21 and 22 events we did find two wavefronts: the primary wavefront was detected during the earlier episode of energy release; a narrower, jet-associated wavefront during the later episode (Figures~\ref{fig:init} and \ref{fig:init0621}, and accompany animations). In the June 25 event, only the primary wavefront was detected; a 2nd wave front, even if existed, could be easily missed because the 2nd SXR peak is weaker compared with the other two flares, and the active region was close to the limb by that time.

\subsection{Wave Interaction with Coronal Structures}

The transmission of the primary wavefront through the bulk of the polar coronal hole CH2 was consistently observed on 2015 June 21 (Figures~\ref{fig:wave0621}e), 22 (Figure~\ref{fig:sector}(Sectors 15--17)) and 25 (Figure~\ref{fig:wave0625}(d--f)), with CH2 being imaged from quite different viewing angles, which minimizes the possibility of false detection. On the other hand, the transmission through the equatorial coronal hole CH1 was only seen at its southern end on June 21 (Figure~\ref{fig:wave0621}d) and northern end on June 22 (Figure~\ref{fig:sector}; Sector 8). We noticed that a bright stationary front formed at the AR-facing boundary of CH1 (Figures~~\ref{fig:prop}, \ref{fig:wave0621} and \ref{fig:wave0625}), sustaining for an extended time period in each case. Hence we speculate that the primary wavefront only transmitted \replaced{trough}{through} the ends of CH1 because the bulk of wave energy was dissipated at the stationary front. Such stationary fronts also appear in numerical experiments when a fast-mode wave passes a coronal hole \citep[e.g.,][]{Piant2017,Piant2018b} or a quasi-separatrix layer \citep{Chen2016}.  On 2015 June 22, a reflected/refracted front was seen to move away from the polar coronal hole at projected heights up to $\sim\,$110 Mm above the limb (Figure~\ref{fig:arc}). We note that a similar height ($\sim\,$80--100 Mm) was obtained for EUV waves using STEREO quadrature observations \citep{Patsourakos2009,Kienreich2009}.     

The propagation of the primary wavefront along a coronal arcade consisting of east-west oriented loops was consistently observed in all three events (Figures~\ref{fig:prop}, \ref{fig:wave0621}f, and \ref{fig:wave0625}(g--i)). When we had a clear view of the whole arcade (June 22 and 25), we found the propagation speed was significantly decelerated as the wavefront propagated at the far-side of the arcade (Figures~\ref{fig:prop}(g--i) and \ref{fig:wave0625}(g--i)); but when the arcade was located close to the east limb  (June 21), its far side was obscured by the perspective, and we found no clear deceleration (Figure~\ref{fig:wave0621}f). Hence the deceleration was most likely a foreshortening effect involving the arched geometry and viewing angles. While the wavefront apparently slowed, it was enhanced in 211~{\AA} and dimmed in 193 and 171~{\AA}. We speculate that the arcade serves as a lens that `focuses' the waves toward the far end of the arcade, resulting in a local heating as evidenced by the different responses in 211 (\ion{Fe}{14}), 193 (\ion{Fe}{12}), and 171~{\AA} (\ion{Fe}{9}).

Under the impact of the primary wavefront from the south, the northward displacement of the filaments was negligible compared to the southward displacement (Figures~\ref{fig:ff} and \ref{fig:ff0621}). This can be understood by the Alfv\'{e}n speed increasing with height in the low corona (below about 3~\rsun) of the quiet Sun \citep{Gopalswamy2001}. As a result, fast magnetosonic wavefront tends to be refracted toward the surface, consequently inclining forward, which was indeed observed in some limb events \citep[e.g.,][]{hudson03, LiuW2012} and reproduced by numerical simulations \citep[e.g.,][]{wu01, grechnev11b}. Upon arriving at a filament suspended in corona, the compressive wavefront would push the filament down forward due to the oblique front. The resultant displacement is expected to be small partly because of increasing magnetic field and plasma density toward the surface, and partly because of projection effects with the filament being located in the northern hemisphere. The filament then swings back and is expected to overshoot because of the rarefaction in the wake of the compressive wavefront, therefore displaying a significant southward displacement. The 2nd swing of the filament towards its original position would again exert pressure on the surface. Such impacts may produce the northward-propagating secondary wavefronts. These wavefronts are very diffuse probably because they originate at different times from different places along the filament. With speeds comparable to the sound speed, these secondary wavefronts are likely slow-mode waves. 

\subsection{Nature of the Primary Wavefront}
The magnetic field in the polar coronal hole is open and hence predominantly vertical to the wave propagation, while in the coronal arcade the magnetic field is traced by coronal loops and hence parallel to the wave propagation. This reveals the nature of the primary wavefront, since only fast magnetosonic waves can propagate in all directions with respect to the magnetic field. On the other hand, the stationary front at the boundary of the equatorial coronal hole makes it unlikely to interpret the primary wavefront as a slow-mode soliton \citep{Wills-Davey2007,Long2017}. Although compatible with the stationary front, both the current shell model  \citep{Delannee2007,Delannee2008} and the continuous reconnection model \citep{Attrill2007} cannot accommodate the transmission through, and the reflection/refraction from, the polar coronal hole \citep{Long2017}. The fast-mode shock wave in the field-line stretching model \citep{Chen2002,Chen2005} is consistent with the primary wavefront in our observations, but it would be difficult for this model to explain why a slower front appeared only when the primary front impacted the filaments, had the slower front resulted from the stretching of field lines overlying the erupting flux rope. 

The 2015 June 25 event is optimal for observing the transmission of the primary wavefront through the polar coronal hole and the coronal arcade (Figure~\ref{fig:wave0625}a), as both structures were close to the disk center. Given that the wave propagates at the fast-magnetosonic speed, in the coronal hole the speed is 
\begin{equation}
v=\sqrt{v_\mathrm{A}^2+c_s^2},
\end{equation}
but in the coronal arcade
\begin{equation}
v=v_\mathrm{A},
\end{equation}
where $v_\mathrm{A} = B/\sqrt{4\pi nm_p}$ is the Alfv\'{e}n speed, and $c_s = \sqrt{\gamma k_BT/m_p}$ is the sound speed. These speeds are insensitive to plasma temperature and density because of the square root. We made a rough estimate of the magnetic field strength by adopting typical values of plasma density number in the corona, i.e., $n\sim 10^8$~cm$^{-3}$ for the coronal hole and $10^9$~cm$^{-3}$ for the coronal arcade. We further assumed that the plasma temperature is in the range of 1--2~MK in the quiet Sun, and that the adiabatic index $\gamma=5/3$. With the wave propagation speeds measured in the time-distance maps (Figures~\ref{fig:wave0625}(d \& g)), we found that $B\sim 2$~G and plasma $\beta\sim0.2$--0.3 in the coronal hole, and that $B\sim4$~G and $\beta\sim0.5$--1 in the coronal arcade neighboring the active region. These numbers are comparable with previous results obtained by various methods \cite[][and references therein]{Liu+Ofman2014}.

\subsection{Summary}
\begin{figure} 
	\centering
	\includegraphics[width=0.85\hsize]{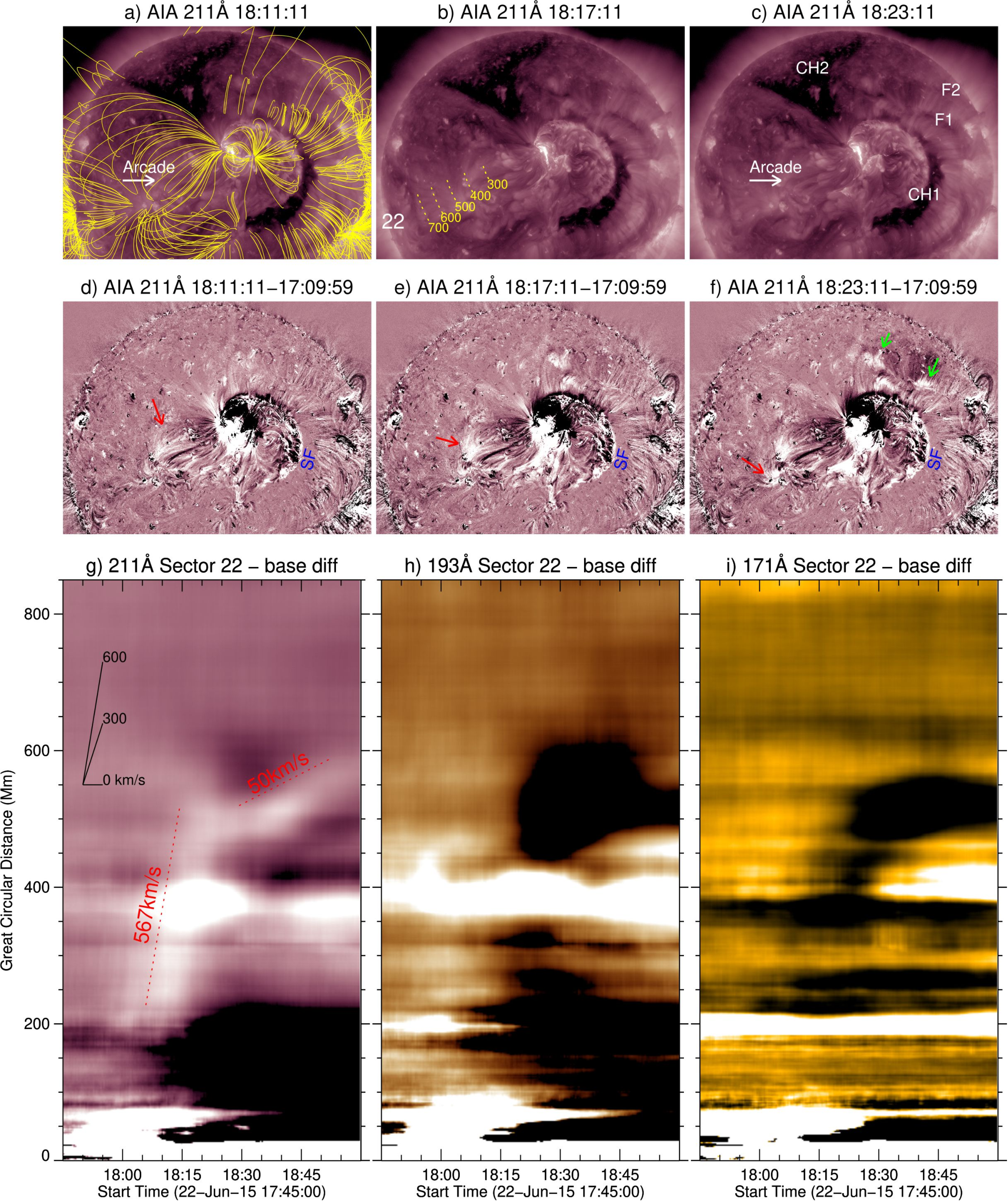}
	\caption{Wave propagation along an arcade of coronal loops on 2015 June 22. (a--c) Snapshots of AIA 211~{\AA} images. Panel (a) is superimposed by closed field lines given by the PFSS model, same as in Figure~\ref{fig:slit}d. Sector \#22 from Figure~\ref{fig:slit}a across the arcade of interest is replotted in Panel (b). Yellow dashed curves indicate the distance from the sector center, with numbers in units of Mm. (d--f) Base different images in AIA 211~{\AA} corresponding to images in Panels (a--c). Red arrows mark the disturbance propagation along the arcade. Green arrows mark the secondary wavefronts induced at the quiescent filament F1. (g--i) time-distance maps constructed via Sector \#22 as indicated in Panel (b), using AIA base difference images in 211, 193, and 171~{\AA}, respectively. An animation of 211~{\AA} base-difference images is available online.  \label{fig:prop}}
\end{figure}

\begin{figure} 
	\centering
	\includegraphics[width=0.9\hsize]{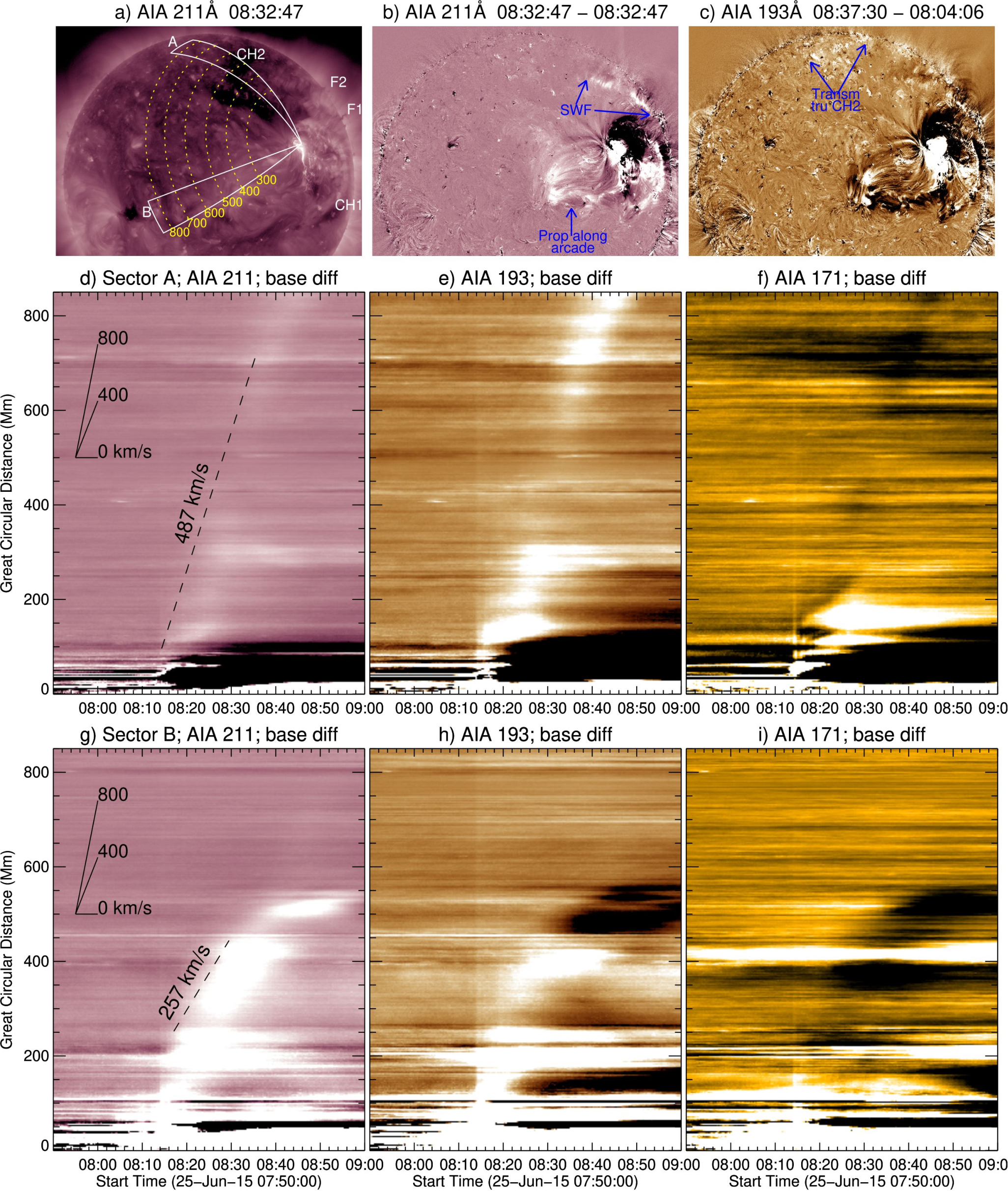}
	\caption{EUV Waves in the 2015 June 25 event. Panel (a) shows two representative sector-shaped slices (labeled `A' and `B'). Yellow dashed circles indicate the distance from the sector center, i.e., the flaring site, with numbers in units of Mm. (b \& c) Running difference images in 211 and 193~{\AA} showing the wave propagation along an arcade of coronal loops to the east of AR 12371, secondary wavefront induced at the filament F1, and wave transmission through the coronal hole CH2. A stationary front is also visible outlining the boundary of the  crescent-shaped coronal hole CH1. Middle (bottom) row shows wave propagation along Sector A (B). The time-distance maps are constructed with base-difference images in three AIA passbands, 211, 193, and 171~{\AA}. An animation of 211 and 193~{\AA} base-difference images is available online. \label{fig:wave0625}}
\end{figure}

We have analyzed the coronal EUV waves associated with the three CMEs on 2015 June 21, 22 and 25, respectively. These homologous eruptions provide us a rare opportunity to investigate how the waves interact with coronal structures from three different viewing angles. Such observations help resolve the ambiguity and confusion due to projection and perspective effects, and yield new insight into the physics of EUV waves, as being highlighted below.
\begin{itemize}
	\item The propagation both along the field in the coronal arcade and across the field in the coronal holes substantiates the primary wavefront as fast magnetoacoustic waves. The wave nature is further corroborated by the reflected/refracted wavefront bent away from the polar coronal hole. 
	\item As the primary wavefront propagates along a coronal arcade toward its far end, a slow front is seen to `veer off' from the fast front in the time-distance maps (Figures~\ref{fig:prop} and \ref{fig:wave0625}). This is an effect of perspective and projection, which we suspect might be responsible for some of the slow fronts exhibiting similar features in time-distance maps \citep[e.g.,][]{Chen+Wu2011,Asai2012,Xue2013,Shen2014wave}. This offers an alternative scenario to those interpreting the slow front as a pseudo-wave produced by coronal restructuring associated with CMEs \citep[e.g.,][]{Delannee2008,Chen2005,Attrill2007}. 
	\item Slow and diffuse wavefronts are produced by the primary wavefronts impacting on filaments, which may account for some extremely slow and diffuse ``EIT'' waves \citep[e.g.,][]{Warmuth+Mann2011}.	
\end{itemize}

\acknowledgments The authors thank the anonymous referee for critical comments and constructive suggestions that help improve the manuscript. R.L. acknowledges support by NSFC 41474151, 41774150, and 41761134088. Y.W. acknowledges support by NSFC 41131065 and 41574165. JL was supported by the NSFC grants 41331068, 11790303 (11790300), and 41774180. This work was also supported by NSFC 41421063, CAS Key Research Program KZZD-EW-01-4, and the fundamental research funds for the central universities.


\clearpage

\end{document}